\documentclass[floatfix,twocolumn]{revtex4}
\pdfoutput=1
\usepackage{graphicx}
\usepackage{amsmath}
\usepackage{latexsym}
\usepackage{color}
\usepackage{epsfig}

\begin{document}

\title{Ultracold dense gas of deeply bound heteronuclear molecules}

\author{S. Ospelkaus,$^1$ A. Pe'er,$^1$ K.-K. Ni,$^1$ J. J. Zirbel,$^1$ B. Neyenhuis,$^1$ S. Kotochigova,$^2$ P. S. Julienne,$^3$ J. Ye,$^1$ and D. S. Jin$^1$ }
\affiliation{$^1$JILA, National Institute of Standards and Technology and University of Colorado,
Department of Physics, University of Colorado, Boulder, CO 80309-0440, USA \\
$^2$Physics Department, Temple University, Philadelphia, PA 19122-6082, USA\\
$^3$Joint Quantum Institute, National Institute of Standards and Technology
and University of Maryland, Gaithersburg, Maryland 20899-8423, USA
}

\begin{abstract}
Recently, the quest for an ultracold and dense ensemble of polar molecules has attracted strong interest. Polar molecules have bright prospects for novel quantum gases with long-range and anisotropic interactions~\cite{anisotropic}, for quantum information science~\cite{quantumcomp}, and for precision measurements~\cite{precision1, precision2, precision3}. However, high-density clouds of ultracold polar molecules have so far not been produced. Here, we report a key step towards this goal. Starting from an ultracold dense gas of heteronuclear $^{40}$K$^{87}$Rb Feshbach molecules~\cite{exp1,exp2} with typical binding energies of a few hundred kHz and a negligible dipole moment, we coherently transfer these molecules into a vibrational level of the ground-state molecular potential bound by $>10\,$GHz. We thereby increase the binding energy and the expected dipole moment of the $^{40}$K$^{87}$Rb molecules by more than four orders of magnitude (see Table~\ref{comparison}) in a single transfer step. Starting with a single initial state prepared with Feshbach association~\cite{Feshbach}, we achieve a transfer efficiency of 84\%. While dipolar effects are not yet observable, the presented technique can be extended to access much more deeply bound vibrational levels and ultimately those exhibiting a significant dipole moment. The preparation of an ultracold quantum gas of polar molecules might therefore come within experimental reach.
\end{abstract}

\maketitle

Strategies for the realization of an ultracold dense gas of polar molecules have generally followed two different approaches. The first is to directly cool ground-state polar molecules by means of buffer gas cooling~\cite{buffergas}, Stark deceleration~\cite{Starkdeceleration1, Starkdeceleration2} or velocity filtering~\cite{velocity}. However, direct cooling strategies have typically been restricted to the mK temperature range. Photoassociation schemes~\cite{photoassociation}, on the other hand, have the advantage of using high-density ultracold atomic clouds as a promising starting point. However, the association of colliding atoms into deeply bound molecular levels is hindered by the poor wave function overlap between the continuum states of colliding atoms and the localized molecular states~\cite{limit1,limit2}. This wavefunction overlap can be significantly enhanced by starting from a weakly bound molecule as compared to the continuum state of two colliding atoms. Weakly bound molecules in well-defined near-dissociation quantum states can be efficiently created in the vicinity of magnetic-field-tunable Feshbach resonances in ultracold atomic gases~\cite{Feshbach}. The resulting ensembles of Feshbach molecules constitute an excellent launching stage for the application of coherent optical transfer schemes to produce molecules in deeply bound vibrational levels. The combination of heteronuclear Feshbach molecule creation~\cite{hetFesh1,hetFesh2, exp2} with coherent deexcitation schemes~\cite{Avi2007} might therefore ultimately result in the creation of an ultracold dense gas of polar molecules.

\begin{figure}
\centering
\includegraphics[width=0.9\columnwidth]{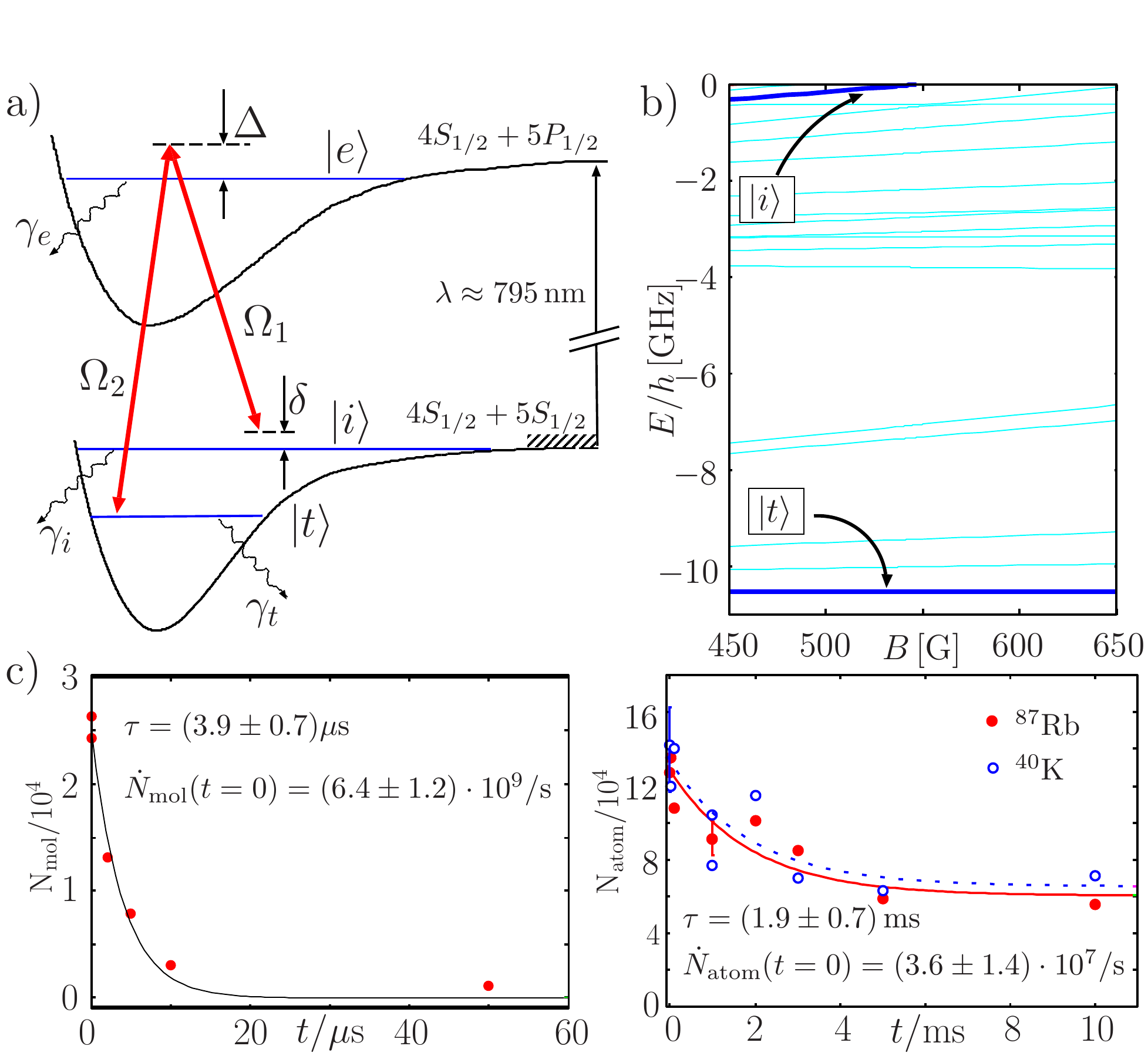}
\caption{(a) Schematic illustration of molecular potentials and molecular vibrational levels involved in the coherent two-photon transfer scheme. The phase-coherent lasers 1 and 2 couple
the initial  $\left|i\right>$ and the target state $\left|t\right>$ to the intermediate state $\left |e\right>$ in the excited-state molecular potential connecting to the $4S_{1/2}+5P_{1/2}$ threshold.
$\Omega_1$ and $\Omega_2$ denote the corresponding Rabi frequencies. $\delta$ and $\Delta$ are the two-photon and single-photon detuning, respectively. $\gamma_x$ represents the decay rate of state $x$. (b) Calculated molecular
energy structure of the $^{40}$K$^{87}$Rb molecule in the ground-state molecular potential below the K
$\left|9/2,-9/2\right>$Rb$\left|1, 1\right>$ atomic threshold. All levels have total spin projection quantum number
$\mathrm{M_F}=-7/2$.  The initial Feshbach state $\left|i\right>$ and the target state $\left |t\right>$ are highlighted by the blue bold lines.  (c) Comparison of  loss rates starting from Feshbach molecules (left panel) to photoassociation rates from free atoms (right panel), respectively. In the right panel, the Rb (K) atom number is plotted with the red solid (open blue) circles. The red solid and blue dashed lines are simultaneous fits to the decaying Rb and K atom numbers. Here, we assume that the loss rate ($\dot{N}_\mathrm{atom}$) is the same for both components at any time.} \label{level_struct}
\end{figure}

In the homonuclear $\mathrm{Rb}_2$ system, Feshbach molecules have been used as a starting point for coherent two-photon transfer into a state bound by $\sim$500\,MHz, which corresponded to increasing the binding energy by a factor of 20~\cite{coherent_transfer}. In addition, rf manipulation schemes have been used in the $\mathrm{Rb}_2$ system achieving binding energies of $>3\,$GHz~\cite{cruising}. These results represent a powerful demonstration of the degree of control
available in ultracold gases. Here, we report on coherent transfer of heteronuclear Feshbach molecules into a deeply bound vibrational level of the ground-state molecular potential. We thus demonstrate a key step towards the realization of a quantum degenerate ultracold gas of \textit{polar} molecules. Using two phase-coherent laser fields to couple the initial Feshbach state $\left|i\right>$ and the target state $\left|t\right>$ to a common electronically excited state $\left|e\right>$ (see Figure~\ref{level_struct} a), \textbf{STI}mulated \textbf{R}aman \textbf{A}diabatic \textbf{P}assage (STIRAP)~\cite{STIRAP} is used to coherently transfer the population of $\left|i\right>$ to $\left|t\right>$. By this transfer process, we increase the binding energy of the heteronuclear molecules by more than 4 orders of magnitude and produce a dense ultracold cloud of heteronuclear molecules with a binding energy of $>10$\,GHz. An essential prerequisite for the successful demonstration of STIRAP in this system has been careful one-photon spectroscopic investigation of the so-far unknown $^{40}$K$^{87}$Rb excited-state molecular potential below the $4S_{1/2}+5P_{1/2}$ threshold 
and precise two-photon spectroscopy of more deeply bound vibrational levels in the ground-state molecular potential of $^{40}$K$^{87}$Rb.

\begin{table}[hbt]
\resizebox*{0.9\columnwidth}{!}{
\begin{tabular}{|l||ccc|}
    \hline
Molecular Property  &  \textrm{Feshbach molecule}    &   $a^3\Sigma^+(v=-3)$  & $X^1\Sigma^+(v=-99)$    \\
    \hline \hline
\textrm{Binding Energy (GHz)}& 0.00027 & 10.49238 & $\approx 125000$  \\
\textrm{Size ($a_0$)} & $270$ & $34$ & $7.7$ \\
\textrm{Dipole moment ($ea_0$)} & $2\times 10^{-11}$ & $4.1\times 10^{-5}$ & $0.3$ \\
\hline
\end{tabular}
} \caption{Comparison of the binding energy, size, and dipole moment of
the Feshbach molecules, the current target state
$a^3\Sigma^+(v=-3)$, and the absolute ground state of the
$X^1\Sigma^+(v=-99)$ potential~\cite{dipole_mom}. $v$ is the vibrational quantum
number as counted from threshold, with the least bound level labeled
as $v=0$. } \label{comparison}
\end{table}

Starting from a near quantum-degenerate Bose-Fermi mixture of $^{87}$Rb and $^{40}$K atoms confined in a single-beam optical dipole trap, we use  magnetic-field ramps across a Feshbach resonance at $546.7\,$G to create $2\times 10^4$ heteronuclear Feshbach molecules  with a density of $\approx 5\cdot10^{11}/\mathrm{cm}^3$. At a magnetic field of $B=545.88$\,G, the Feshbach molecules have a binding energy of $(270\pm 50)$~kHz and can be directly imaged by high-field resonant absorption imaging. To suppress inelastic collisions of the molecules with remaining unbound atoms, we remove about 95\% of the left-over atoms. At the chosen binding energy of $270$~kHz, the Feshbach molecules are in a superposition of the open channel K$\left|F=9/2,m_F=-9/2\right>+$Rb$\left|F=1,m_F=1\right>$  and the adiabatically connecting closed channel K$\left|7/2,-7/2\right>+$Rb$\left|1,0\right>$~\cite{exp2}. Here $F$ denotes the total atomic spin and $m_F$ the spin projection along the magnetic-field direction.

The Feshbach molecules are manipulated with light derived from a phase-coherent Raman laser system.  A Ti:Sapphire laser is offset locked to a temperature-stabilized Fabry-Perot cavity, resulting in a linewidth of $<$20\,kHz and absolute long-term frequency stability better than 2 MHz. The second laser beam is derived from an external cavity diode laser phase-locked to the Ti:Sapphire laser. The two beams are $\pi$-polarized with respect to the magnetic field and propagate collinearly when irradiated onto the molecular ensemble.

In the first step, we perform one-photon, bound-bound spectroscopy of the electronically excited $^{40}\mathrm{K}^{87}\mathrm{Rb}^*$ molecular potentials below the $4S_{1/2}+5P_{1/2}$ threshold. Starting from the Feshbach molecules with a binding energy of $270$\,kHz at $B=545.88$\,G, laser 1 is used to couple the molecules to vibrational levels in the excited-state molecular potentials. These electronically excited molecules subsequently decay into deeply bound vibrational levels of the ground-state molecular potential where they are invisible to our detection method. Using loss spectra, we identify vibrational series of different excited-state molecular potentials, which will be reported elsewhere.  After an extensive search, we have chosen the $2(0^+)\,\, v=-14$ state as an intermediate coupling state $\left|e\right>$. Here $v$ denotes the vibrational quantum number as counted from the $4S_{1/2}+5P_{1/2}$ threshold. This state is located $213$\, GHz below the $4S_{1/2}+5P_{1/2}$ threshold.  Comparing one-photon  loss rates starting from Feshbach molecules to photoassociation rates of free atoms, we observe an enhancement in the excitation rate by  $(180\pm60)$ (see Figure~\ref{level_struct}\,c).

\begin{figure}[tbp]
\begin{centering}
\leavevmode
\resizebox*{0.9\columnwidth}{!}{\includegraphics{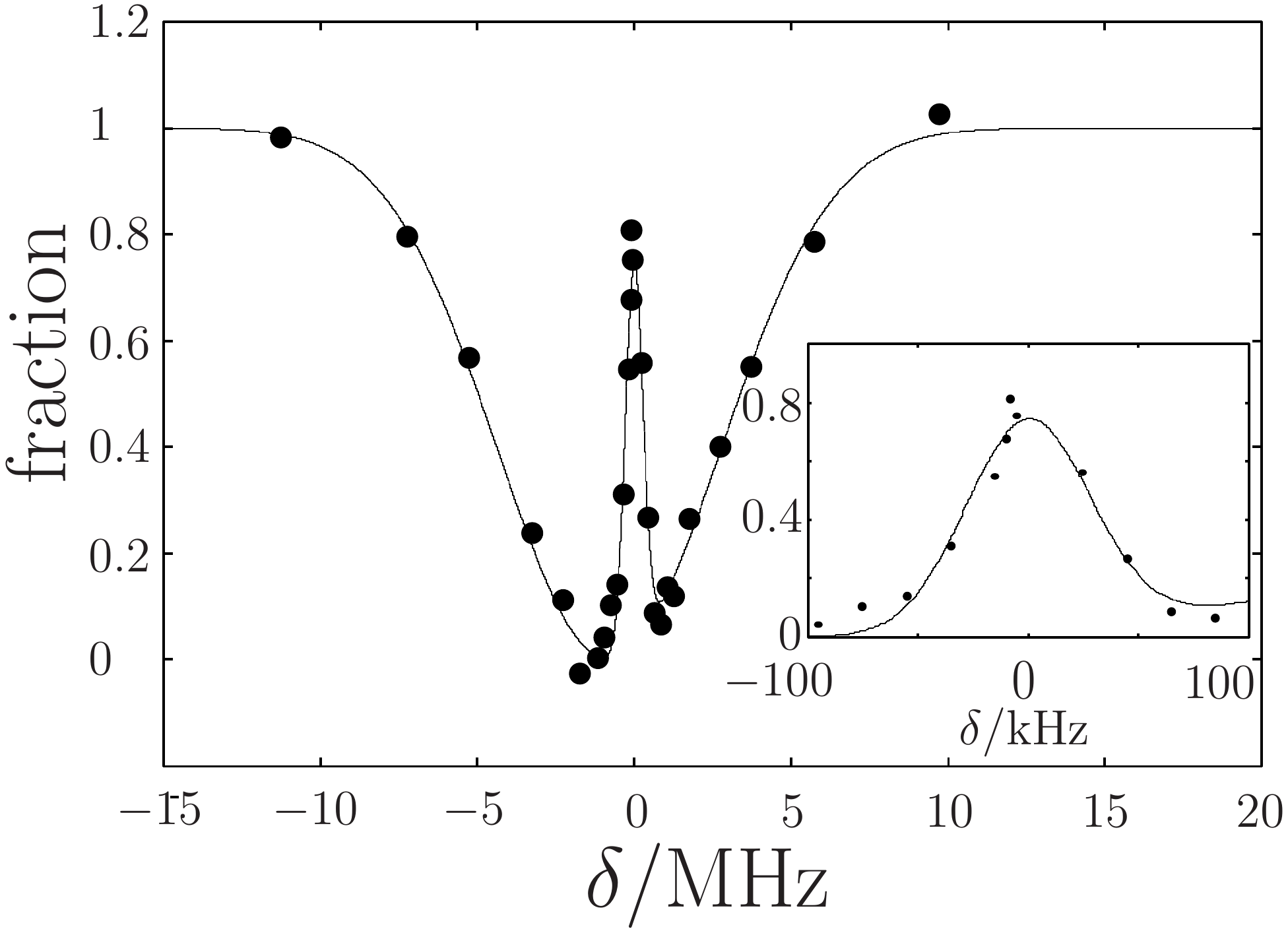}}
\end{centering}
\caption{ Dark resonance in the molecular system. In this case, the target state $\left|t\right>$ is the K$\left|7/2,-7/2\right>+$Rb$\left|1,0\right>\,\,v=-3$ state. We shine a $10\,\mu$s square-pulse of Raman light onto the Feshbach molecules and the remaining fraction of Feshbach molecules is detected. For $\delta=0$, the loss of Feshbach molecules is strongly suppressed due to the emergence of a dark state. For this particular set of data, $\Delta\approx -2\pi\cdot 0.8$\,MHz. The inset shows the central dark resonance feature. }
\label{darkres}
\end{figure}

An essential prerequisite for  STIRAP is  precise knowledge of possible target states. While a theoretical model of the electronic ground-state potential has recently been published by Pashov
\textit{et al.}~\cite{potentialKRb}, the ground-state level structure of $^{40}$K $^{87}$Rb in the near-threshold regime has never been probed experimentally. Using two-photon dark-resonance spectroscopy~\cite{dark_res}, we probe the binding energy of vibrational levels in the K$\left|9/2,-9/2\right>+$Rb$\left|1,1\right>$  and K$\left|7/2,-7/2\right>+$Rb$\left|1,0\right>$ channels at $B=545.88$\,G. We probe levels with binding energies less than $10.5$\,GHz, which is accessible to our phase-locked laser system.  Figure~\ref{darkres} shows a typical dark resonance spectrum when scanning the frequency difference ($\delta$) of the two phase-coherent laser fields in the vicinity of the frequency splitting between the two molecular ground-state levels. In the limit $\gamma_e\gg\Omega_2\gg\Omega_1$, the transition $\left|t\right>\rightarrow\left|e\right>$ (see Figure~\ref{level_struct}) is dressed by the near-resonant laser 2, leading to destructive interference for the absorption probability of laser 1 on the probe transition $\left|\mathrm{i}\right>\rightarrow\left|\mathrm{e}\right>$ for $\delta=0$. We observe one-photon loss of Feshbach molecules with a width of 6\,MHz and a narrow dark resonance feature in
the vicinity of $\delta=0$. At $B=(545.88\pm 0.05)$\,G, we measure a binding energy of $3.1504(10)$\,GHz and $10.49238(15)$\,GHz for the K$\left|9/2,-9/2\right>+$Rb$\left|1,1\right>$ $v=-2$ and $v=-3$ states, respectively. For
K$\left|7/2,-7/2\right>+$Rb$\left|1,0\right>\,v=-3$, we find $7.31452(15)$\,GHz. The measured binding energies deviate by less than $0.1$\,\% from  the theoretical prediction based on the
potentials by Pashov \textit{et al.}~\cite{potentialKRb}.

\begin{figure}[tbp]
\begin{centering}
\leavevmode
\resizebox*{0.9\columnwidth}{!}{\includegraphics{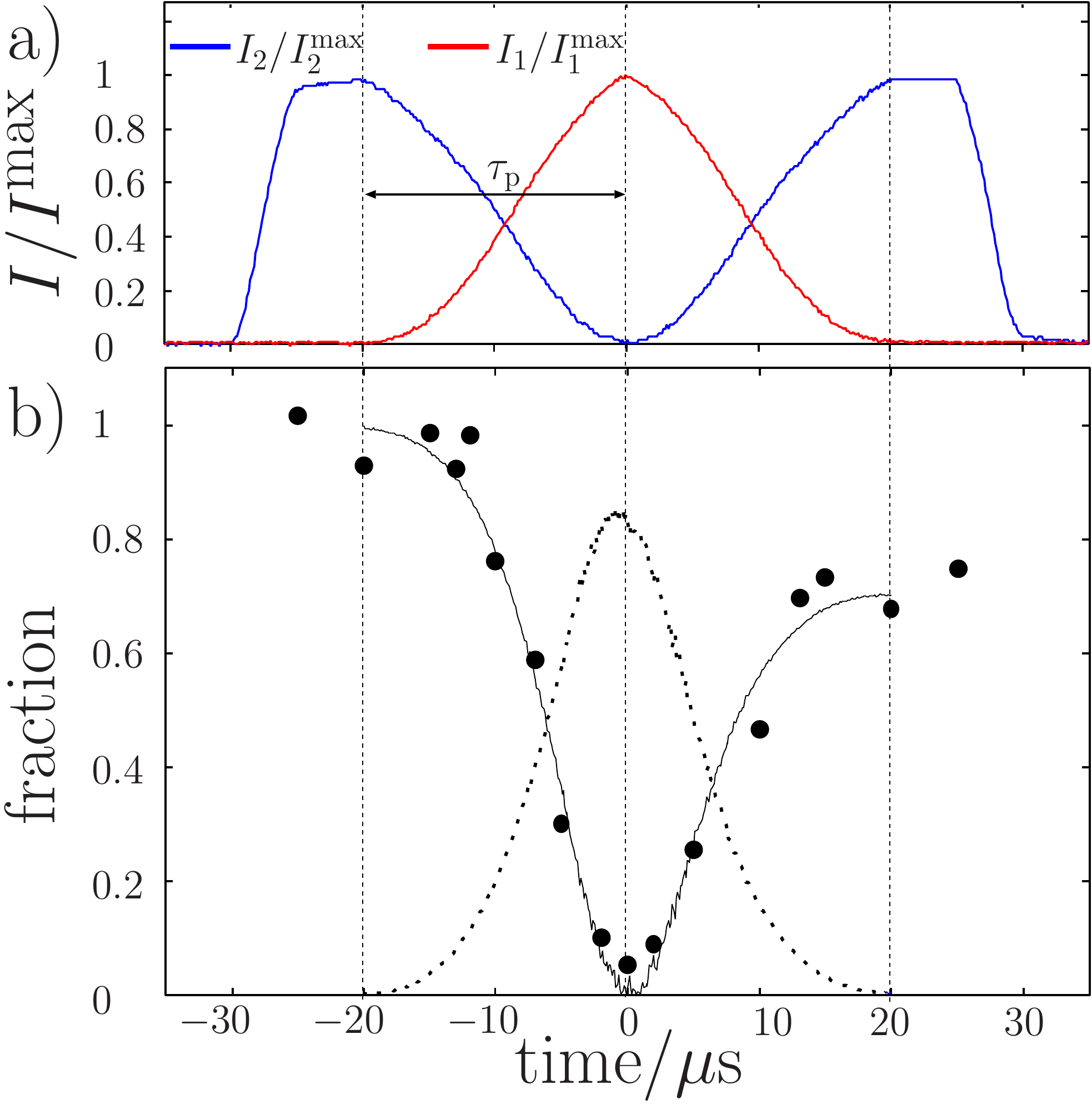}}
\end{centering}
\caption{ Time evolution of the coherent two-photon transfer. (a) Counterintuitive STIRAP pulse sequence. (b)  Measured population (black circles) in the initial Feshbach state $\left|i\right>$  during the STIRAP pulse sequence. Starting with a pure cloud of Feshbach molecules (population of 1 in $\left|i\right>$), the Feshbach molecules are coherently transfered to the target state $\left|t\right>$ by the first pulse sequence ($t=-\tau_{\mathrm{p}}$ to $t=0$). In the deeply bound target state, the molecules are invisible to the detection light. Reversing the pulse sequence, the molecules are transfered back to the initial state $\left|i\right>$ ($t=0$ to $t=\tau_{\mathrm{p}}$). The solid line (dashed line) is a theoretical calculation of the input state  (target state) population based on eq. (\ref{eq1}). }
\label{timeseries}
\end{figure}

With possible target states and coupling states precisely determined, we use STIRAP to transfer the molecules
coherently into the 10.49238\,GHz bound vibrational level of the ground-state molecular potential. STIRAP is known to be a  robust coherent transfer scheme in a three-level system. When two-photon resonance ($\delta=0$) is maintained, the molecules are transfered between $\left| i\right>$ and $\left|t\right> $ with negligible
population in the lossy excited state $\left|e\right>$ throughout the process. Figure~\ref{timeseries}\,a) shows the counterintuitive STIRAP pulse sequence used in the experiment. In the first step, laser 2 is turned on, coupling the target state $\left|t\right>$ to the intermediate state $\left|e\right>$. While the intensity of laser 2 is ramped down from $I^{\mathrm{max}}_2$ to 0 within  $\tau_p=20\,\mu$s, the intensity of laser 1 is ramped up to its maximum value $I^{\mathrm{max}}_1$, thereby adiabatically transferring the population from the Feshbach state into the deeply bound target state $\left|t\right>$. Reversing the pulse sequence reverses the process and the transfer occurs from the deeply bound level $\left|t\right>$ to the initial state $\left|i\right>$, as shown for $t>0$ in Fig.~\ref{timeseries}\,a).

Figure~\ref{timeseries}\,b) shows the time-dependence of the measured population of the initial Feshbach state $\left|i\right>$ during the pulse sequence.  Molecules in the Feshbach state can be detected by
direct high-field resonant absorption imaging, whereas molecular population in the target state is invisible to the light.  During the pulse sequence, we observe the hiding of the molecules in the more deeply bound vibrational level and the transfer back into the initial state after reversal of the pulse sequence. We observe an efficiency of the double STIRAP sequence of $71$\%, corresponding to an efficiency of $84$\% for a single pulse~\cite{life_eff}. The transfer process can be described by an open three-level system in the Rotating Wave Approximation:
\begin{equation}
i\cdot \left( \begin{array}{c} \dot{\Psi}_i(t)\\ \dot{\Psi}_e(t) \\ \dot{\Psi}_f(t)
   \end{array}\right)
 = \left( \begin{array}{ccc}
\delta-\frac{i \gamma_i}{2}  & \frac{\Omega_1\left(t\right)}{2} & 0\\
        \frac{\Omega_1\left(t\right)}{2} & \Delta-\frac{i\gamma_e}{2} & \frac{\Omega_2(t)}{2}\\
    0 & \frac{\Omega_2(t)}{2} & \frac{i\gamma_f}{2}. \end{array} \right)
\cdot \left( \begin{array}{c} \Psi_i(t)\\ \Psi_e(t) \\ \Psi_f(t)
   \end{array}\right),
\label{eq1}
\end{equation}
Here, $\left|\left<\Psi_x(t)|\Psi_x(t)\right>\right|^2$ and $\gamma_x$ are the population and decay rate of state $x$, respectively (see Fig.~\ref{level_struct}).  The time-dependence of the transfer is determined by the exact pulse shape and the ratio of the maximum Rabi frequencies $\Omega^{\mathrm{max}}_1$ and $\Omega^{\mathrm{max}}_2$. Given the pulse shape shown in Fig.~\ref{timeseries}\,a),  we can extract the Rabi frequency ratio
used in the experiment. Given $I^{\mathrm{max}}_2/I^{\mathrm{max}}_1=0.27$, we obtain
$\Omega^{\mathrm{max}}_2/\Omega^{\mathrm{max}}_1=(1.7\pm0.1)$, corresponding to a ratio of the effective transition dipole moments of $(3.3\pm0.3)$.

\begin{figure}[tbp]
\begin{centering}
\leavevmode
\resizebox*{1.0\columnwidth}{!}{\includegraphics{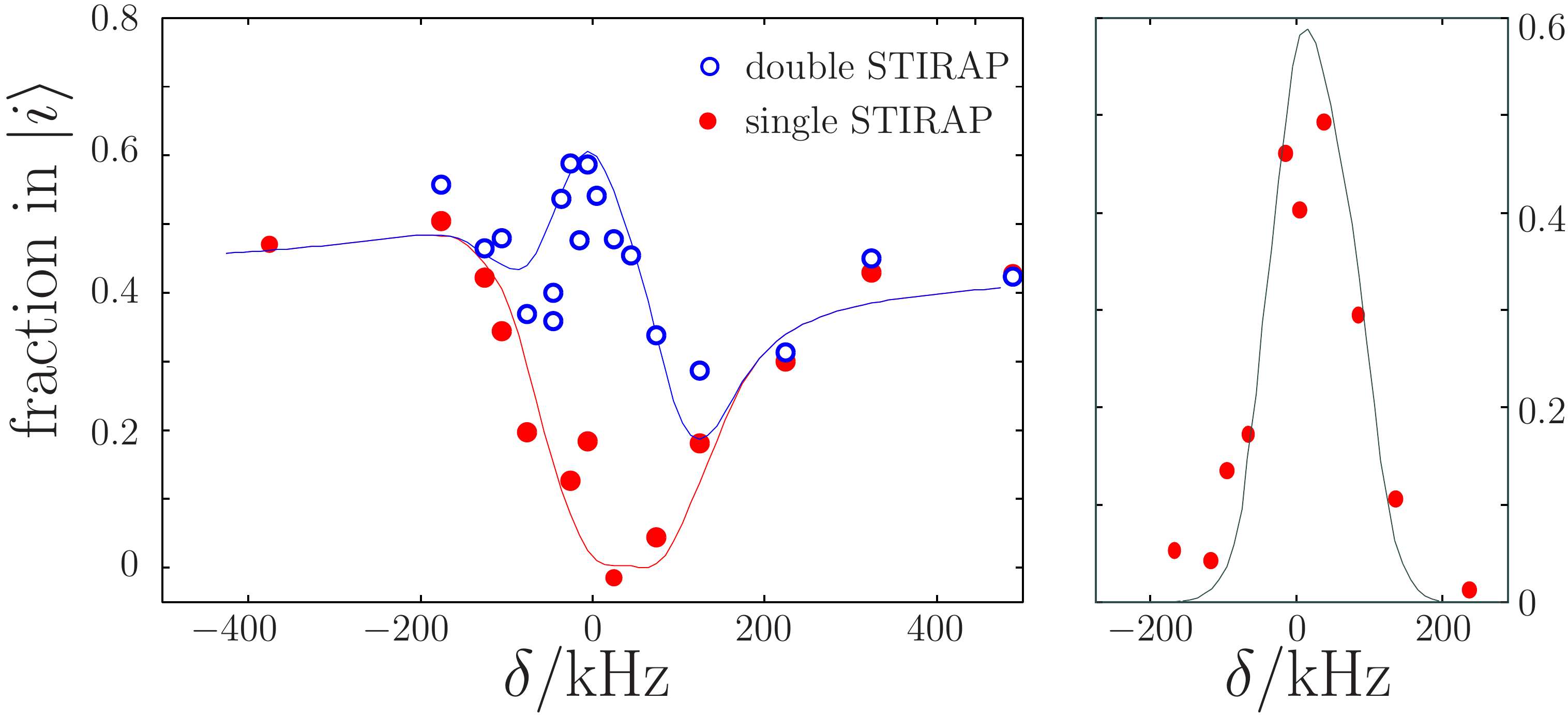}}
\end{centering}
\caption{ STIRAP lineshape (a) STIRAP lineshape for $\Delta=-2\pi\cdot 50\,$MHz, $\Omega_1=2\pi\cdot 3.6$\,MHz, $\Omega_2/\Omega_1=1.7$ both for single STIRAP (transfer from $\left|i\right>$ to $\left|t\right>$) and double STIRAP (transfer from $\left|i\right>$ to $\left|t\right>$ and back). The asymmetry in the lineshape is due to the finite single-photon detuning $\Delta$ (see text). (b) Difference between the double and single STIRAP signal as plotted in (a). The difference corresponds to the fraction of Feshbach molecules transfered into the deeply bound state and back at a given $\delta$.
}\label{lineshape}
\end{figure}

The absolute coupling strengths can be extracted from an analysis of the STIRAP efficiency as a function of the two-photon detuning $\delta$, as shown in Fig.~\ref{lineshape}. We have measured the lineshape both for a single STIRAP process ($\left|i\right>$ $\rightarrow$ $\left|t\right>$) and for a double STIRAP ($\left|i\right>$ $\rightarrow$ $\left|t\right>$ $\rightarrow$ $\left|i\right>$). A comparison between a simulation based on
Eq.~\ref{eq1} with $\Omega_2/\Omega_1=1.7\pm 0.1$ and $\Delta=-2\pi\cdot (50\pm5)$\,MHz and the experimental single STIRAP lineshape data yields $\Omega^{\mathrm{max}}_1=2\pi\cdot (3.6\pm 0.5)$\,MHz and $\Omega^{\mathrm{max}}_2=2\pi\cdot (6.1\pm 0.9)\,$MHz at $I^{\mathrm{max}}_1=\left(3.7\pm1.5\right)\,\mathrm{W/cm^2}$ and  $I^{\mathrm{max}}_2\left(1.0\pm0.4\right)\,\mathrm{W/cm^2}$. This translates into effective transition dipole moments of $d^{\mathrm{eff}}_1=(0.050\pm0.015)\,e a_0$ and $d^{\mathrm{eff}}_2=(0.17\pm0.04)\,ea_0$.

\begin{figure}[tbp]
\begin{centering}
\leavevmode
\resizebox*{0.8\columnwidth}{!}{\includegraphics{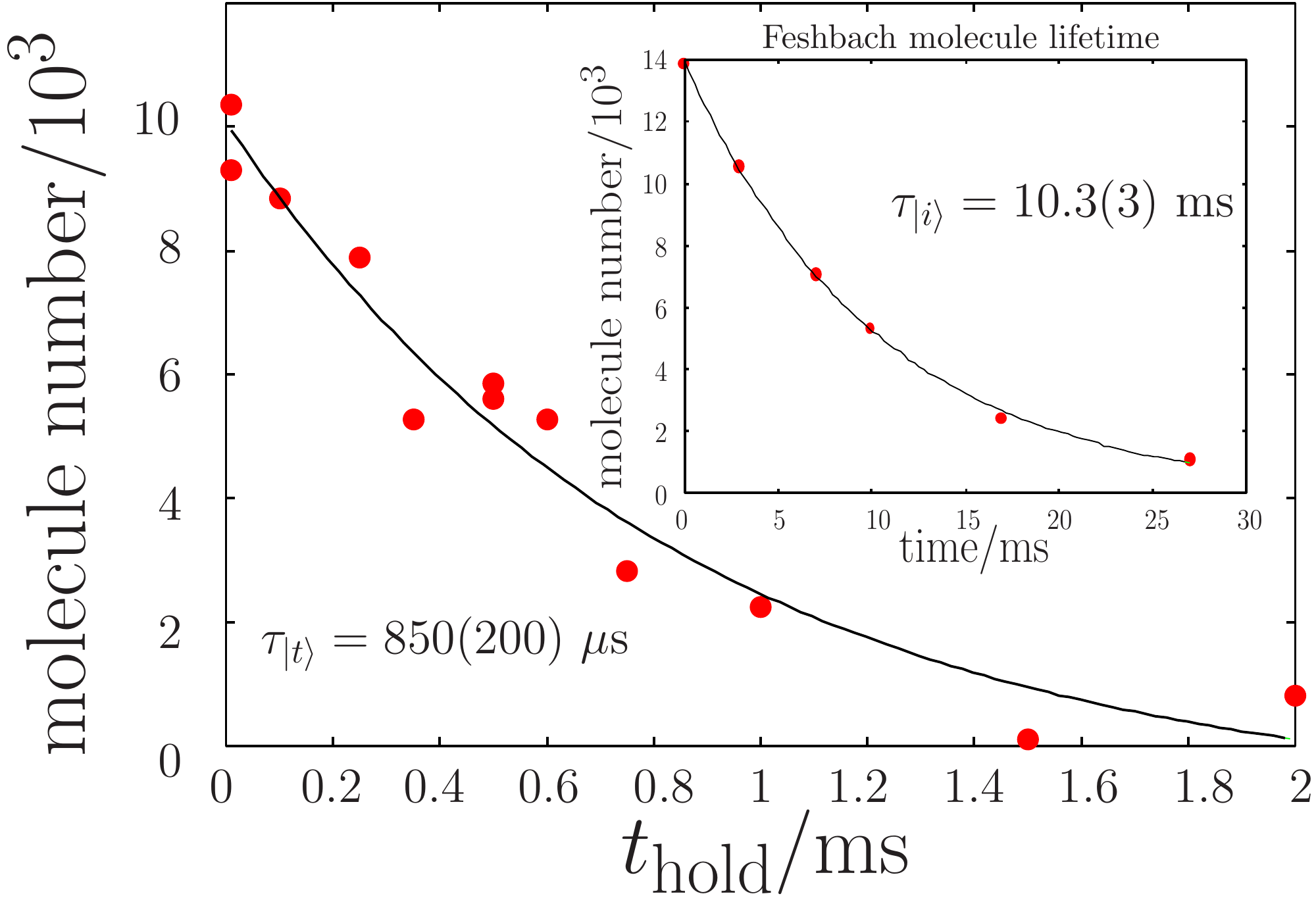}}
\end{centering}
\caption{ Lifetime of 10.49238(15)\,GHz bound molecules. We observe a lifetime of $850\,\mu$s. The inset shows the decay of the initial Feshbach molecules, which have a lifetime of $10.3$\,ms.
}\label{decay}
\end{figure}

Finally, we have measured the lifetime of the deeply bound molecules in the target state. Starting from the molecules in state $\left| i\right>$, we apply a single STIRAP pulse to transfer the molecules into the target state $\left| t \right>$. The deeply bound molecules are held in the optical dipole trap for varying times $t_{\mathrm{hold}}$. A second STIRAP pulse then transfers the molecules back to the initial state where they can be detected by high-field resonant absorption imaging. Figure~\ref{decay} shows the decay of the deeply bound molecules. We observe  a lifetime of the molecules in $\left| t\right>$ of $(850\pm 200)\,\mu$s as compared to a lifetime of the initial state of $(10.3\pm 0.3)$\,ms. We observe the lifetime of both Feshbach molecules and deeply bound molecules decrease with increasing atom density. This suggests that atom-molecule collisions limit the molecule lifetime.

In conclusion, we have demonstrated coherent optical transfer of heteronuclear Feshbach molecules into a 10\,GHz bound vibrational level of the ground state molecular potential. The presented techniques can easily be extended to binding energies $>10\,$GHz by referencing both lasers to a frequency comb and therefore being able to bridge almost arbitrary frequency gaps~\cite{CundiffYe}. Our experiments show that the binding energy of the molecules can be enhanced by more than four orders of magnitude in a single STIRAP step. Starting from these $10\,$GHz bound molecules, another four orders of magnitude will be necessary to access the absolute rovibrational ground state (see Table~\ref{comparison}). The production of a polar molecular quantum gas by means of coherent optical transfer of heteronuclear Feshbach molecules might hence come within experimental reach.

\begin{acknowledgments}
We acknowledge funding support from NIST, NSF, and DOE. K.-K. N. and B. N. acknowledge support from the NSF,
S. O. from the Alexander-von-Humboldt Foundation and P. S. J. from the ONR. We thank D. Wang for stimulating discussions and C. Ospelkaus for critical reading of the manuscript.
\end{acknowledgments}

\appendix

\subsection*{Feshbach molecule formation}

\noindent{After evaporation of $^{40}$K and $^{87}$Rb in an Ioffe-Pritchard magnetic trap, $5\times10^6$ $^{87}$Rb and $1\times10^6$ $^{40}$K atoms are loaded into a single-beam optical dipole trap operating at $1064\,$nm. At this stage, the temperature of the atoms is about $3\,\mu$K.  In the optical trap, rf adiabatic rapid passage is used to prepare a $^{40}$K$\left|9/2,-9/2\right>$ and $^{87}$Rb$\left|1,1\right>$ spin mixture. It is in this combination of Zeeman states that the Feshbach resonance  at a magnetic field of $546.7$\,G occurs between K and Rb. After ramping the magnetic field to $555\,$G, we further evaporate the mixture in the optical dipole trap to a temperature of $T/T_c\approx 1$. Here, $T_c$ is the critical temperature for the onset of Bose-Einstein condensation of the Rb cloud. Starting from the near-degenerate mixture of $1\times10^5$ $^{40}$K atoms and $3\times10^5$ $^{87}$Rb atoms, we finally use  a magnetic-field ramp  across the $546.7\,$G Feshbach resonance to couple the open scattering channel to the bound molecular state, thereby associating about $2\times 10^4$ heteronuclear Feshbach molecules. The molecules have a  temperature of $300\,$nK and  a density of $\approx 5\cdot10^{11}/\mathrm{cm}^3$. At a magnetic field of $B=545.88$\,G, the binding energy of the molecules is $(270\pm 50)$~kHz.} 

\subsection*{Direct resonant absorption imaging of heteronuclear Feshbach molecules}

\noindent{The weakly bound heteronuclear Feshbach molecules (binding energy of 270\,kHz)  are detected by direct absorption imaging. The imaging uses light resonant with a cycling transition starting from the  $^{40}$K$\left|9/2, -9/2\right>$ state. For small binding energies of the heteronuclear molecules, the atomic cycling transition of $^{40}$K is detuned only by a few MHz from the molecular transition. Light resonant with the atomic transition will therefore  dissociate the molecules and then scatter on the resulting K atoms. To distinguish the absorption signal of weakly bound Feshbach molecules from unbound $^{40}$K atoms, we clean out residual $^{40}$K$\left|9/2, -9/2\right>$  atoms with an  rf $\pi$-pulse on the atomic $\left|9/2,-9/2\right>\rightarrow\left|9/2,-7/2\right>$ transition prior to imaging.
Note that direct imaging of molecules at high-magnetic field is unique for \textit{heteronuclear} Feshbach molecules.  In the heteronuclear case, both the ground and the excited electronic potentials share  a $1/R^6$ long-range dependence. The difference in energy between these two states therefore varies more slowly with the internuclear separation than in the homonuclear case where the excited state long-range potential varies as $1/R^3$.}


\begin{thebibliography}{22}
\bibitem{life_eff}
Given the lifetime of the deeply bound state of $850\,\mu$s, we would ideally expect an efficiency of 83\% for the double STIRAP. The additional loss is presumably due to residual population of the lossy excited state during the process.
\end{thebibliography}
\end{document}